# Uniform intensity in multifocal microscopy using a spatial light modulator

M. Junaid Amin,[1,2,3,4] Sabine Petry,[1] Haw Yang [2] and Joshua W. Shaevitz[3,4,*]

[1]*Department of Molecular Biology, Princeton University, Princeton, New Jersey 08544, USA*
[2]*Department of Chemistry, Princeton University, Princeton, New Jersey 08544, USA*
[3]*Lewis-Sigler Institute for Integrative Genomics, Princeton University, Princeton, New Jersey 08544, USA*
[4]*Department of Physics, Princeton University, Princeton, New Jersey 08544, USA*
*\*shaevitz@princeton.edu*

**Abstract:** Multifocal microscopy (MFM) offers high-speed three-dimensional imaging through the simultaneous image capture from multiple focal planes. Conventional MFM systems use a fabricated grating in the emission path for a single emission wavelength band and one set of focal plane separations. While a Spatial Light Modulator (SLM) can add more flexibility, the relatively small number of pixels in the SLM chip, cross-talk between the pixels, and aberrations in the imaging system can produce non-uniform intensity in the different axially separated image planes. We present an in situ iterative SLM calibration algorithm that overcomes these optical- and hardware-related limitations to deliver near-uniform intensity across all focal planes. Using immobilized gold nanoparticles under darkfield illumination, we demonstrate superior intensity evenness compared to current methods. We also demonstrate applicability across emission wavelengths, axial plane separations, imaging modalities, SLM settings, and different SLM manufacturers. Therefore, our microscope design and algorithms provide an alternative to fabricated gratings in MFM, as they are relatively simple and could find broad applications in the wider research community.

## 1. Introduction

Multifocal microscopy is a useful method that allows simultaneous imaging of multiple object planes to realize high-speed 3D imaging. Common implementations of multifocal microscopy use a fixed, fabricated phase mask optimized for a specific wavelength or object-plane separation distance [1-3]. The fabricated grating can be replaced with a Liquid Crystal Spatial Light Modulator (SLM) that can dynamically change the phase mask for use with multiple wavelengths or object plane separations [4-5]. However, uniform illumination across the subimages is difficult to achieve using an SLM-based phase mask due to inherent device characteristics including pixel-to-pixel crosstalk effects [6-7]. We present an in situ iterative calibration method for the generation of optimized SLM phase patterns that produce multifocal images with near-uniform subimage brightness.

## 2. Prior-art method for obtaining uniform subimage brightness in multifocal microscopes

The phase grating in a multifocal microscope has two tasks: (i) to divide incoming emission light equally into a 2D array of orders, and (ii) to axially offset the orders to different object planes separated by a distance $\Delta z$ by modifying the phase pattern using a geometric distortion function. The Pixelflipper algorithm was designed to generate uniformly-illuminated subimages in existing multifocal microscope systems [2]. This algorithm uses an in situ iterative procedure that finds the phase pattern of a grating unit cell of size $P_u \times P_u$ pixels$^2$ that gives the highest uniformity among the diffraction orders in the computed Fourier plane. This optimized unit cell is then repetitively arranged into a grid to provide the phase pattern to be displayed on the SLM. Pixelflipper often fails to produce adequate results using SLMs as it assumes an aberration free optical system that takes the Fourier transform precisely. Furthermore, in addition to other system aberrations, SLMs suffer from pixel-to-pixel crosstalk effects [6-7] that further alter the resultant diffraction pattern. The same issue occurs when iterative Fourier transform algorithms [8] are used, particularly when only few SLM pixels form the repeated grating pattern. These effects are illustrated in the image of a 100 nm Gold Nanoparticles (AuNPs) under darkfield illumination on an SLM-based multifocal microscope (see Appendix) using $\Delta z = 0$ nm (Fig. 1(a)). There is an undesirably significant difference in subimage intensities in the image even though the Pixelflipper algorithm was used to optimize the SLM pattern with $P_u = 4$ (Fig. 1(b)).

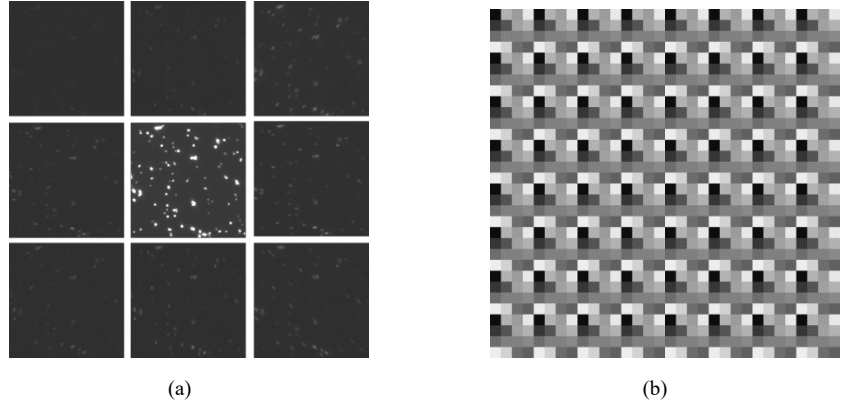

(a)          (b)

Fig. 1. (a) In-focus sub-images cropped from a 9 plane z stack acquired from the camera resulting from deploying a Pixelflipper algorithm [11] optimized SLM display pattern. The emission light is unevenly distributed in the subimages, with the central zeroth order subimage receiving most of the emission light, and (b) zoomed-in view of a 64 × 64 pixels$^2$ region of the SLM displayed grating pattern which gives the multifocal subimages in (a).

## 3. In situ iterative calibration routine for subimage intensity uniformity

Here, we propose an in situ iterative calibration method to generate phase patterns which allows near-uniform illumination in the subimages of an SLM-based multifocal microscope. The algorithm is based on a feedback loop between the SLM, camera, and the computer, and uses real-time images from the camera to update the pattern on the SLM until the optimal grating pattern is acquired. To evaluate the subimage intensity uniformity for our optimization routine, we modified the metric proposed in ref. [9] and used instead,

$$M = \frac{\min(\{I_{m,i}\}) - I_b}{\max(\{I_{m,i}\}) - I_b}, \qquad (1)$$

where $\{I_{m,i}\}$ is the measured subimage $i$ ($i = 1 \ldots N \times N$) and $I_b$ is a measured background intensity. $M$ ranges from 0 to 1, with $M=1$ corresponding to completely uniform subimage intensities. We initially generate 100 random unit cells and start with the one that has the highest $M$ value. The unit cell is then updated iteratively by sequentially iterating over all graylevel values for all pixels in the unit cell. This routine is repeated until there is no change in the unit cell during a complete iteration over all pixel locations.

This calibration method gives visually uniform subimage intensities using $\Delta z = 0$ nm (Fig. 2(a)), when displaying an in situ iterative optimized pattern on the SLM (Fig. 2(b)). The computed $M$ value for this pattern is 0.712, much larger than the $M$ value measured when using the Pixelflipper-based phase mask ($M = 0.033$, Fig. 1(a)). We compared multiple trials of the

Pixelflipper algorithm, our in situ iterative method, and randomly-generated phase patterns using $P_u = 4$ (Fig. 3). The in situ iterative method shows superior performance over both the Pixelflipper and the randomized methods, realizing multifocal images having large $M$ values, i.e., near uniform subimage intensities (Fig. 3).

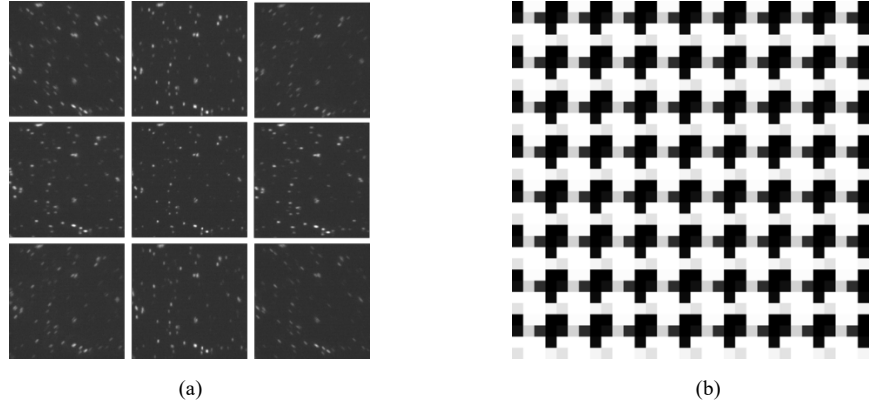

(a)                                                                 (b)

Fig. 2. (a) In-focus sub-images cropped from a 9 plane z stack acquired resulting from deploying our in situ iteratively optimized SLM display pattern. The emission light striking the camera is more evenly distributed among the subimages compared to Fig. 1(a), and (b) zoomed-in view of a $64 \times 64$ pixels$^2$ region of the SLM displayed grating pattern which gives the images in (a).

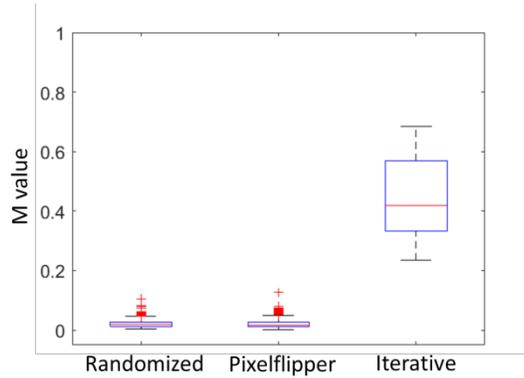

Fig. 3. Boxplots of output M values resulting from grating patterns optimized using Pixflipper (500 iterations, Randomized (500 iterations) and our in situ iterative calibration algorithm (12 iterations).

## 4.  Multifocal Imaging of Biological Specimen

We used our in situ iteratively optimized SLM phase patterns to acquire 3D images of GFP-labeled tubulin in MeOH-fixed TPX2 Hela Kyoto cells [10]. The in situ iterative calibration algorithm was first executed using darkfield imaging under the current settings to obtain the optimized SLM phase pattern, which was then phase distorted using the algorithm in [2] to achieve object plane separation in the subimages. Multifocal snapshots of the sample cells using 488-nm laser excitation are obtained (Fig. 4). The sequence of the subimages in the multifocal image (Fig. 4) follows the sequence shown in Appendix Fig. A1(b), with the top right plane of the image corresponding to the $z = -4\Delta z$ plane, where $\Delta z$ represents the focal plane separation in object space. The $\Delta z$ values for the images (Fig. 4) are 0.50 µm (Fig. 4(a)) and 1.00 µm (Fig. 4(b)). The different focal cross-sections of the cells can be visibly seen across the subimages, displaying the structural features in the sample and the 3D imaging power of the multifocal microscope.

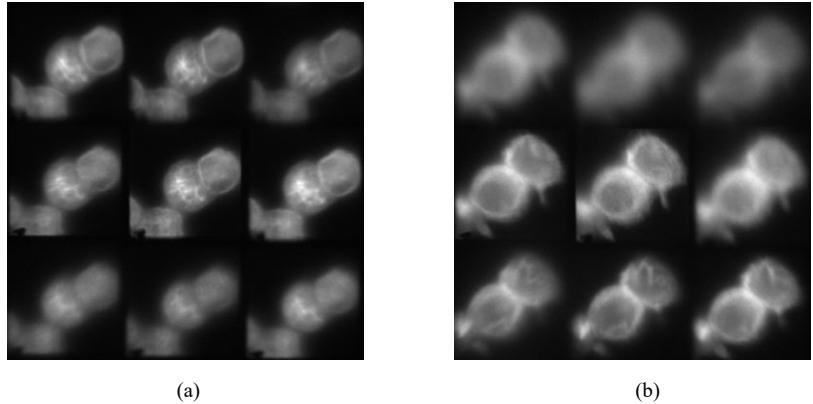

(a) (b)

Fig. 4. Multifocal Images of different regions of a microtubule stained Hela Kyoto Cells fixed sample with (a) Δz = 0.50 μm, and (b) Δz = 1.00 μm.

## 5. Discussion

We demonstrate that our in situ iterative optimization algorithm is effective at generating SLM-based patterns which allow uniformity across multifocal subimages. The key underlying concept of the algorithm is the inclusion of real-time experimental variables into the optimization process. Many of the parameters involved such as $P_u$, the emission wavelength and the SLM model are empirical hardware choices. In particular, the $P_u = 4$ value throughout the paper is empirically chosen to maximize the imaging field of view without overlapping of the subimages in the current system optical settings. Other microscope setups may require a different unit cell size to be optimal.

The method allows comparable MFM imaging performance when using SLMs versus fabricated gratings. Deploying SLMs as multifocal gratings has numerous advantages: they are available off the shelf, require no additional investment other than its initial cost, and can readily be programmed to change any multifocal grating parameter including Δz values as well as number of simultaneous imaging planes at high speed, limited by the refresh rate of the SLM (typically 60 Hz). Fabricated gratings, on the other hand, require access to clean room facilities having fabrication and lithography tools which require extensive training to use. This process is expensive, time-consuming and not readily available near many research labs. Using our in situ iterative algorithm, more researchers can now build SLM based MFMs for investigating fast microscopic 3D processes in biology, physical chemistry and other domains. The universal applicability of the calibration routine is demonstrated to account for different SLM manufacturers, wavelength, unit cell sizes as well as different microscope modalities (see Appendix), making this method widely applicable to the different imaging requirements. Future work involves exploring various other optimization techniques and metrics to further improve subimage intensity uniformity in the SLM based MFM.

## 6. Appendix

*6.1. SLM based Multifocal Microscope (SLM-MF)*

*6.1.1. Optical Diagram*

Fig. A1(a) illustrates the optical design of the home-built multifocal microscope. Three main segments of the system are highlighted for easier visualization: Darkfield-Brightfield (DF-BF) illumination, laser illumination and multifocal Imaging. In the DF-BF illumination module, incoherent light from a white LED passes through a Diffuser before being focused by a lens L1 onto an iris. Lens L2 then collimates the light and sends it onto the DF Mirror to realize Darkfield Imaging mode. The DF Mirror is a custom designed mirror with an oval central transmissive area thereby allowing reflection of the incoming light into a ring-shaped intensity pattern downwards into the condenser objective. To switch from Darkfield mode to Brightfield imaging mode, the DF Mirror can be replaced by the BF Mirror, where the BF Mirror is a conventional fully reflecting optical mirror. Note that when operating in Darkfield mode, the Detection Objective's Numerical Aperture (NA) is kept smaller than that of the Condenser Objective.

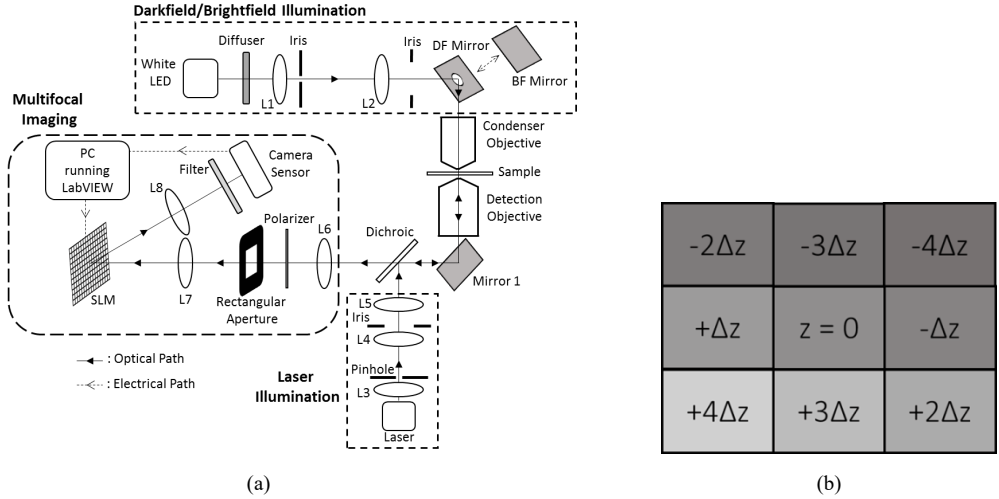

Fig. A1. (a) Optical diagram of the SLM based Multifocal Microscope, and (b) front view of the camera sensor showing the placement of the simultaneously sub-images, each corresponding to a unique object plane.

In the laser illumination module, a laser is spatially filtered and collimated by a combination of lens L3, a Pinhole and lens L4. Lens L5 is introduced to focus the beam reflecting off a dichroic to the back focal plane (BFP) of the detection objective for epi-illumination. This beam will be used for excitation of fluorescent samples. Note that although fluorescence samples can also be used for the experimental calibration routine, they aren't optimal due to bleaching induce changing intensities in the images relative to more stable darkfield and brightfield imaging modes.

In the multifocal imaging module, lens L6 functions as a tube lens and forms an image of the sample at the plane of the rectangular aperture through a linear polarizer. The rectangular aperture's function is to control the imaging Field of View (FOV), thereby preventing the obtained subimages from overlapping at the imaging plane downstream. The polarizer is necessary for the phase-only function of the deployed SLM. Lens L7, having a focal length $f_7$, is placed a focal length's distance from the rectangular aperture, while the reflective SLM is located at the Fourier plane of L7. With the focal lengths of L6 and L7 the same, the SLM is essentially conjugate to the BFP of the detection objective which is also the Fourier plane of the sample.

The SLM acts as a multifocal grating in this setup. Each of the diffraction orders emanating from the SLM displayed multifocus grating pattern is encoded with a unique defocus phase (except the unaffected zeroth order). These orders when imaged onto the camera sensor, by Lens L8 (having focal length $f_8$) via an emission filter, form a multifocus image. This image contains N x N subimages with each subimage represents a unique object plane. Details of the algorithm used to generate such grating patterns are described later in the paper. The bandwidth of the emission filter is restricted to around ~15 nm to reduce chromatic dispersion effects originating from the SLM displayed grating. This emission wavelength band restriction can be overcome using chromatic correction optics as used in [2], but is not implemented here. Fig. A1(b) illustration shows the front view of the camera in the multifocus microscope designed for a 3 x 3 array of subimages, though in theory any N x N number of orders can be obtained. Different segments of the camera in Fig. A1(b) correspond to different object planes, with each subimage separated by a distance Δz in object space. Note that the zeroth order remains undiffracted and unaffected by the SLM displayed pattern, and thus corresponds to the z = 0 plane. A Personal Computer (PC) running LabVIEW Software interfaces with the SLM and the camera sensor.

For the SLM-MF, it useful to know the effective lateral FOV as a function of different microscope parameters involved. FOV in this paper is defined as the field of view for each subimage in sample space. To find the expression for FOV, denote Mag as the combined magnification of the detection objective and lens system, $P_u$ the SLM displayed grating period in pixels units, S the SLM pixel size, the C camera pixel size and $\lambda_{min}$ the minimum wavelength in the emission band. The angle θ between the zeroth and 1$^{st}$ orders of a grating is found using the grating equation: $\theta = \sin^{-1}[\lambda_{min}/(P_u \times S)]$. Once θ is known, the FOV can be equated by finding the distance between the centers of both zeroth and 1$^{st}$ orders on the image plane, before dividing by the Mag:

$$FOV = \left(\frac{1}{Mag}\right) f_8 \tan(\theta), \quad (2)$$

### 6.1.2. Experimental Setup

The SLM-MF microscope is custom built in the lab to test the effectiveness of grating patterns designed to optimally distribute incoming light equally into the diffraction orders. For the Darkfield imaging mode, Thorlabs Solis-3C High-Power LED is deployed as the white light LED, along with the accompanying DC20 driver module for intensity control. The diffuser used is Thorlabs DG20-1500. Lens L1 is Thorlabs LA1401-A (focal length = 60 mm) and L2 is AC508-150-A-ML (focal length 150 mm). The DF mirror is custom designed to match the dimensions of the condenser objective used which is the MPLAN BD 50x NA 0.75 objective from Olympus. The detection objective is a Leica 100x, NA 1.4 - 0.7, where the NA is set to 0.7 during Darkfield imaging. For the sample, a mixture is formed using 10 uL of stock 100 nm Gold nanoparticles (AuNPs) solution from BBI solutions and 10 uL of 1M NaCl solution. 10uL is ejected onto a 22 × 22 mm$^2$ coverslip (Fisher scientific) using a pipette before being covered directly by an 18 × 18 mm$^2$ coverslip. The salt is added to immobilize the AuNP onto the coverslip surface. Nail Polish from Electron Microscopy Sciences is used to seal the coverslip edges to avoid leakage of the solution. The sample is mounted onto a P-611.3S NanoCube XYZ Piezo stage from Physik Instrumente using custom machined mounts. Both the piezo stage and the detection objective are mounted on a custom designed aluminum block, which forms the microscopy body. The condenser objective is screwed to a Newport 460A-XYZ translation stage via a custom aluminum adapter plate, with this stage mounted on an 8 inch high post to position the condenser above the sample. Lens L6 is Thorlabs AC508-200-A-ML (focal length = 200 mm), while both $f_7$ and $f_8$ are set to 200 mm (Thorlabs AC254-200-A-ML). The polarizer model is LPVISE100-A, the rectangular aperture is model # 61-1137 from Ealing Catalog, USA, while the emission filter deployed is Semrock FF01-685/10-25 with a central wavelength of 685 nm and FWHM ~ 15 nm. For the biological experiment, the emission filter is changed to one with a 510 nm center wavelength having a bandwidth of ~16 nm. The camera sensor is Hamamatsu's Orca-Flash4.0 V3 sCMOS with a pixel resolution of 2048 x 2048 pixels and a pixel size of 6.5 μm. The camera exposure time is set to 250 ms for the duration of the experiment, and all images are stored as raw 16-bit ".tif" format.

The SLM deployed is the reflective Holoeye PLUTO-2-VIS-056 Phase-only spatial light modulator. It has an 8-bit pixel display resolution, a 1920 x 1080 array of pixels and a pixel pitch S = 8 µm. It can be calibrated, i.e., mapping of displayed gray-level images to actual phase imparted to the incoming light, using different methods. One method is to use the manufacturer provided calibration data specified for certain wavelengths which can be loaded directly onto the SLM via its USB port. Another method involves an experimental procedure outlined in the manufacturer's manual which involves striking the two halves of the SLM with circular top-hat beams originating from the same laser and observing the resulting interference pattern. This latter method is not easily applicable in multifocal setups since the SLM is used in the emission path with emission wavelengths over the visible range. Acquiring lasers for each desired emission wavelength is expensive and impractical, with no guarantee of being effective for a spread of wavelengths as is the case here. Therefore, for this experiment, a default manufacturer-provided calibration curve meant for a 2.2π phase cycle corresponding to 0 – 255 graylevel values for 633 nm laser is uploaded to the SLM firmware. This gives an approximately 2π phase cycle for 696 nm, closely matching the current emission wavelength centered at 685 nm. Apart from the calibration, an important step is to ensure that all SLM displayed patterns have an aperture similar to that of the BFP of the detection objective. Thus, all displayed patterns are multiplied by an aperture function with the central region size corresponding to the size of the BFP. Furthermore, the region outside this aperture is set to a tilted grating with an empirically chosen defocus pattern to steer any stray incoming emission light striking outside the SLM main aperture area away from the zeroth order. Therefore, any subsequent grating patterns will only be displayed inside the central region of the aperture function.

With the experimental setup arranged as described, we implement the prior-art Pixelflipper algorithm described in Section 3.1 using LabVIEW software. N is chosen to be equal to 3, giving a target matrix T of size 3 × 3. With 256 SLM displayable graylevels at our disposal, we limited the graylevel resolution to 80 steps spanning the 0–255 range to reduce the algorithm run time which empirically provides similar performance to having 256 graylevel steps. The Pixelflipper algorithm is run by setting $P_u$ = 4 to give an optimized matrix unit cell U of dimensions 4 × 4. This U is then arranged in a grating format and is phase 'distorted' to give a Δz value of 0.90 µm to realize multifocus imaging, before being multiplied by the aperture function described earlier. Once this resulting pattern is displayed on the SLM, a 9 plane z stack (i.e., images of the sample obtained at multiple z positions by vertical motion of the piezo stage) of the sample is obtained, where the axial spacing of the z stack is chosen to match the deployed Δz = 0.90 µm.

To find the value of M due to a given grating pattern using Eqn. (1), the following procedure is followed: a particle of interest in the field of view is selected and an 80 × 80 pixels² area around that particle is chosen as the Region of Interest (RoI). For each subimage $i$, where $i$ is an integer between 1 and 9, the mean intensity $I_{m,i}$ of its respective RoI is calculated from the image when it is in focus. For example, the mean intensity of subimage 1, $I_{m,1}$, is found by processing only the plane of the z stack when subimage 1 is in focus. Whereas, the mean intensity of, e.g., subimage 7, $I_{m,7}$, is found by processing only that plane of the z stack when subimage 7 is in focus. Once the mean intensities $\{I_{m,i}\}$ of all subimages are found, the minimum and maximum $\{I_{m,i}\}$ values are selected for use in Eqn. (1). To calculate $I_b$, a uniform graylevel pattern is displayed on the SLM which results in the SLM not directing light into the subimages, except into subimage 5 which is the zeroth order and receives most of the incoming light. In this setting, the mean intensities calculated over the same ROIs for all subimages other than subimage 5 are calculated and denoted as $\{I_{b,j}\}$ where $j$ is an integer between 1 and 9, other than 5. $I_b$ is then defined as the minimum value among $I_{b,j}$. In this step, note that the minimum of $I_{b,j}$ is chosen as $I_b$, and not the average of $I_{b,j}$, to avoid negative M values which can occur when $\min(\{I_{m,i}\}) - I_b$ is negative where $I_b$ is brighter than $\min(\{I_{m,i}\})$). Therefore, with $\min(\{I_{m,i}\})$, $\max(\{I_{m,i}\})$ and $I_b$ at hand, Eqn. (1) is used to compute M.

*6.2. SLM-MF subimage field flatness*

The SLM-MF simultaneously images multiple object planes into, e.g., 3 × 3 array of subimages. In addition to optimizing the intensity distribution among the subimages, it is also beneficial to characterize the field flatness of the subimages as a result of the phase distortion implemented in the SLM grating pattern to achieve the 3D imaging capability. An understanding of the uniformity across each subimage is a necessary step in interpreting and processing multifocal 3D imaging data acquired from this microscope.

The following procedure is deployed to characterize the field flatness across the subimages in Darkfield imaging mode: A solution of 200 nm fluorescent beads (660/680) from Life Technologies Corporation is diluted an empirically chosen 400 times before being mixed with a solution of 1 Molar NaCl in a 1:1 volume ratio. 10 µL of this mixture is ejected on a coverslip (22 × 22 mm² coverslip from Fisher scientific) using a pipette before being covered with another coverslip (18 × 18 mm² coverslip (Fisher scientific)). The edges of the smaller coverslip are sealed with nail polish (Electron Microscopy Sciences). As before, the salt helps to immobilize the fluorescent beads to the coverslip surface. The sample is mounted on the piezo stage and a 647 nm Cobolt 130 mW CW laser is used for illumination in an epi-configuration. On the emission side, a ~3nm emission window is created by inserting both Semrock filters FF01-685/10-25 and FF01-685/LP-25 into the light path. The ~3 nm bandwidth significantly minimizes the chromatic dispersion in the subimages (other than the zeroth order subimage 5 which is unaffected). This prevents dispersion related aberrations from negatively affecting the bead localization process described shortly. For illustration,

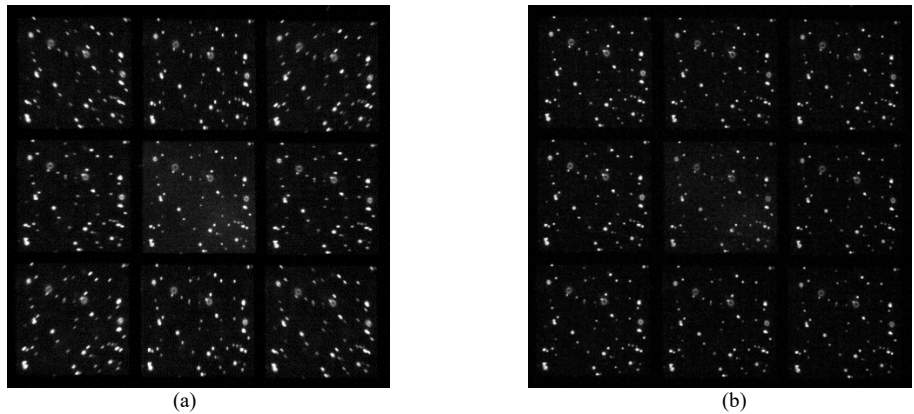

Fig. A2. (a) Images of 100 nm immobilized AuNPs acquired under Darkfield Illumination using emission filter bandwidths of (a) ~ 15 nm, and (b) ~ 3 nm.

multifocal images acquired using ~3 nm and ~15 nm emission filter bandwidths is shown in Fig. A2. Z-stacks are acquired using an in situ iteratively optimized pattern displayed on the SLM having Δz values of 0 µm and 1.00 µm, using stage stapes of 20 nm. For each z-stack, the locations of the beads within the field of view are identified in the lateral and axial Cartesian coordinates. The lateral (*xy*) positions are identified by identifying the bright regions in a focal projection of the z-stacks using the *imfindcircles()* function in MATLAB. Appropriate radii and intensity thresholds are applied to remove possible bead aggregates. The axial (*z*) location of the beads are identified by computing the maximum of the Brenner gradient in a square 16 ×

16 pixels$^2$ region around each identified bead for all images in the z-stack. This localizes the beads in the axial direction.

Once the lateral and axial positions of the beads are found for each subimage, they are plotted in 3D and the *xz* views are displayed in Fig. A3. Fig. A3(a) plot shows displays field flatness data acquired using Δz = 0 µm, whereas Fig. A3(b) corresponds to data acquired when Δz = 1.00 µm. These plots demonstrate near uniform fields across all subimages. Even without correcting for possible sample tilt inherent to the setup, the average peak-valley (P-V) value among subimages acquired using Δz = 0 µm is 0.665 µm while the mean P-V value for Δz = 1.00 µm is 0.467 µm, both well within 2% variation across the field of view signifying reasonably flat subimages most practical multifocal imaging purposes.

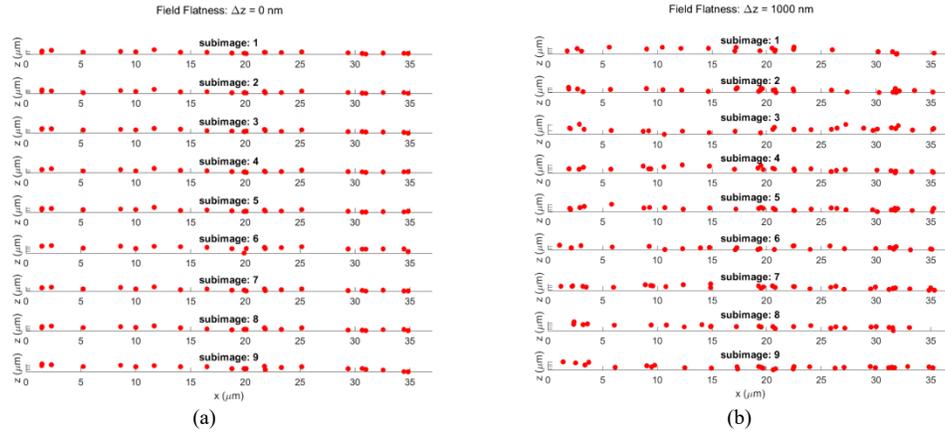

Fig. A3. xz view of the xyz localization of 200 nm beads immobilized on a coverslip to demonstrate the field flatness of the 9 subimages for Δz values of (a) 0 µm, and (b) 1.00 µm.

*6.3. Uniform illumination of orders using Brightfield Imaging mode*

To engage the Brightfield imaging mode, the DF Mirror in Fig. A1(a) is replaced by the BF Mirror to allow full reflection of the light incoming from L2. In this arrangement, we first remove the sample completely, and allow the unscattered light focused by the condenser passes through the detection objective and towards the multifocal optics. Another possibility is to have a cut-out piece of A4 paper as the sample to act as a scattering object. Both methods are tested to work adequately. Apart from the Imaging mode, other key parameter changes for this demonstration includes setting Δz = 0 µm. Since Δz = 0, there is no need to take z-stacks for M value calculation, therefore a single image is obtained for each pattern and the sub-images are processed from it. In addition, M values calculated using the Brightfield imaging mode are denoted $M_{BF}$, to discriminate from M calculated in Darkfield imaging mode. Another parameter change involves setting $f_8$ to 100 mm, which has no effect on the order illumination distribution characteristics but changes the magnification Mag by half.

This arrangement is first tested for a pattern generated by the Pixelflipper using the unchanged $P_u$ = 4 and G = 80. An image acquired using this Pixelflipper optimized pattern is shown in Fig. A4(a). A purple box is shown to annotate the camera region covered spanning the 3 × 3 orders. Fig. A4(b) shows 64 × 64 pixels$^2$ region in the SLM displayed Pixelflipper optimized patter. Visually, the illumination spread among the orders, similar to Fig. 1(a) in terms of contrast, is far from uniform. The computed $M_{BF}$ value for the Fig. A4(a) image is 0.126.

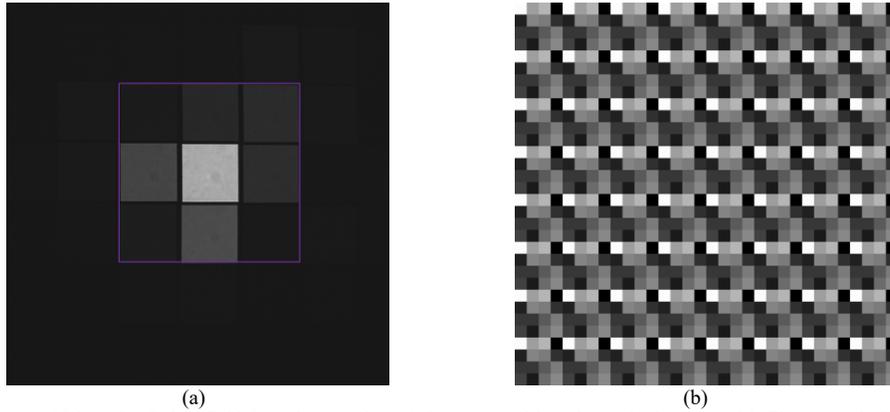
Fig. A4. Using the Brightfield Imaging mode, (a) image resulting from deploying a Pixflipper optimized SLM displayed pattern. The pattern is intended to give a 3 x3 array of uniformly illuminated subimages. The MBF value for (a) is 0.126, and (b) zoomed-in view of the SLM displayed grating pattern which gives the image in (a), showing the repetitive arrangement of unit cells.

Fig. A5(a) shows an output image due to an in situ iteratively optimized pattern, with the purple box annotating the relevant 3 × 3 sub-images region. A 64 × 64 region of the pattern is shown in Fig. A5(b). Visually, Fig. A5(a) shows a higher uniformity of illumination across the orders. The $M_{BF}$ value for Fig. A5(a) is computed to be 0.712. In terms of statistics, the Pixelflipper and randomized pattern generation algorithm are repeatedly executed 1000 times each, and $M_{BF}$ values are computed for each pattern. These $M_{BF}$ values are displayed as a boxplot in Fig. A6 which also shows $M_{BF}$ values from 30 iterations of our algorithm. This plot demonstrates the high degree of illumination uniformity improvement due to our algorithm.

Note that, in Fig. A5(a), additional subimages apart from the bright 3 × 3 subimage array are also visible. These are additional orders which receive illumination from the grating pattern. In the current algorithm framework, only the intensities of the 3 × 3 subimages are optimized, with no correction for the intensity spilling out into the other orders. In future work, new optimization will be explored which allow high efficiency illumination of the subimages while suppressing the unused diffraction orders' intensity to a minimum.

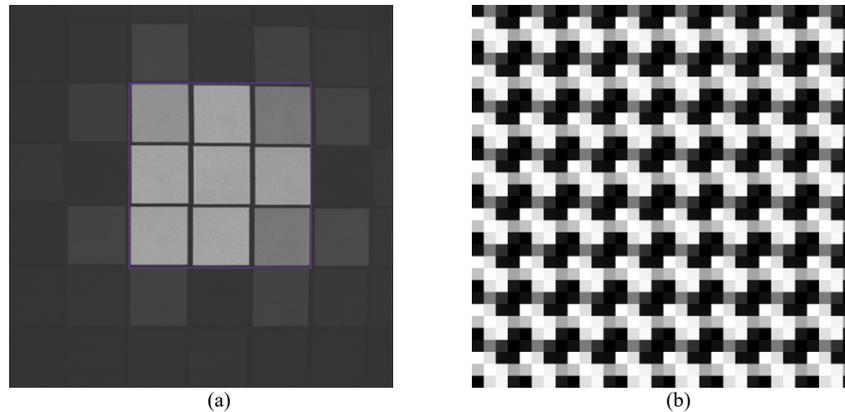
Fig. A5. (a) Image resulting from deploying an output optimized pattern from the proposed in situ iterative algorithm. The pattern is intended to give a 3 x3 array of uniformly illuminated subimages. The M value for (a) is 0.712, and (b) zoomed-in view of SLM displayed grating pattern which gives the image in (a), showing the repetitive arrangement of unit cells.

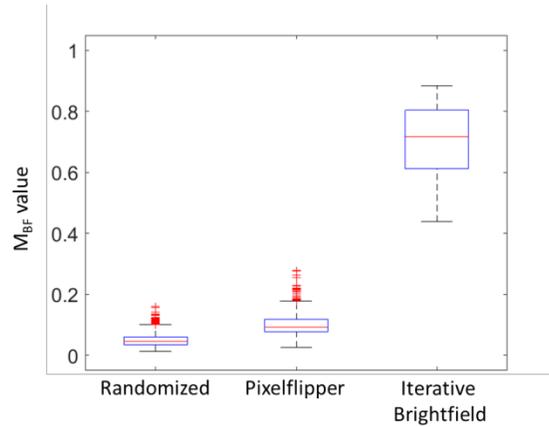

Fig. A6. (a) Boxplots of output MBF values resulting from grating patterns optimized using Pixflipper (1000 iterations), Randomized (1000 iterations) and our in situ iterative algorithm (30 iterations).

### 6.4. Comparison of in situ iterative calibration routine in Darkfield imaging mode versus Brightfield imaging mode

To evaluate the performance of $P_u = 4$ patterns optimized using the in situ iterative calibration method in Brightfield mode, denoted iterative brightfield, compared to the $P_u = 4$ in situ iteratively optimized patterns in Darkfield imaging mode, denoted as iterative darkfield, the 30 brightfield optimized patterns are implemented in Darkfield imaging of the same sample used in obtaining the M value data for Fig. 3 (main text). In this demonstration, all the Brightfield patterns are distorted with $\Delta z = 0.90$ μm. The computed M values resulting from using the Brightfield patterns (30 iterations) in this Darkfield imaging mode are compiled into the Fig. A7 boxplot, which also shows the Fig. 3 (main text) M value data (12 iterations) obtained using our algorithm. This plot indicates that both methods are equally effective in optimizing illumination uniformity across multifocal images.

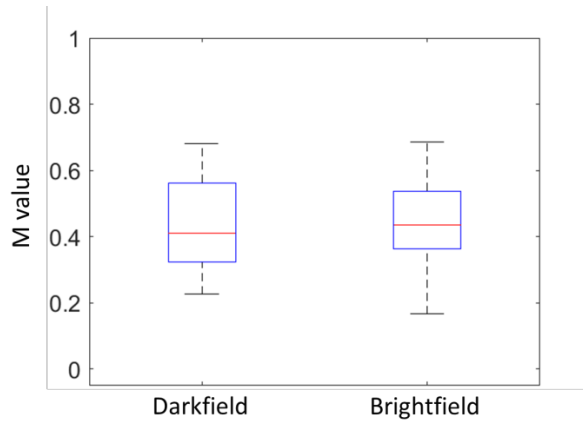

Fig. A7. The boxplot shows a comparison of M values due to 12 different in situ iteratively optimized Darkfield patterns (optimized on AuNP samples under Darkfield imaging) and 30 optimized Brightfield patterns (optimized using Brightfield illumination with no sample) when implemented on the same AuNP sample with $\Delta z = 0.90$ μm.

### 6.5. Pixelflipper output using $P_u > 4$

Our calibration method is demonstrated in this paper to significantly improve the intensity distribution in the multi-focus subimages. Prior to this method, the Pixelflipper has been widely applied to fabricated gratings, though with orders of magnitude larger $P_u$ values. Due to the large pixel sizes of SLMs, larger $P_u$ values equate to large grating periods, which in turn significantly limit the field of view. Selected Pixelflipper algorithm outputs for $P_u$ values of 4, 16 and 32 are shown in Fig. A8, which is demonstrated using the Brightfield imaging mode and using $f_8 = 200$ mm. Additionally, G is set to 256 for all generated Pixelflipper patterns in this demonstration. Fig. A8(a) shows a camera image captured using $P_u = 4$ using the Pixelflipper optimized pattern whose zoomed in 64 × 64 pixels$^2$ region is shown in Fig. A8(b). The FOV for $P_u = 4$ is calculated to be 42.38 μm using $f_8 = 200$ mm, Mag = 100, $\lambda_{min} = 678$ nm and S = 8 μm. Fig. A8(c) shows a camera image captured using $P_u = 16$, with the zoomed in 64 × 64 pixels$^2$ region of the corresponding Pixelflipper optimized pattern shown in Fig. A8(d). In this $P_u$ setting of 16 which realizes larger grating periods displayed on the SLM, the diffraction angle is decreased and the FOV decreases by a factor of 4 to 10.60 μm. In between Fig. A8(a) and Fig. A8(c), the Rectangular Aperture is adjusted to prevent overlap between the subimages on the camera sensor resulting from the $P_u$ increase. Next, $P_u$ is set to 32 and a Pixelflipper optimized pattern is obtained. Fig. A8(e) shows the resulting camera image captured, while Fig. A8(f) shows the zoomed in 64 × 64 pixels$^2$ region of the Pixelflipper optimized pattern. As before, the Rectangular Aperture is adjusted to prevent overlap between the subimages on the camera sensor resulting from the $P_u$ increase from 16 to 32. The FOV in this $P_u = 32$ setting is now 5.30 μm. Qualitatively, according to Fig. A8, the intensity uniformity among the 9 subimages does improve by increasing $P_u$ from 4 to 32. However, this comes at a high cost of eightfold decrease in the FOV, making $P_u = 32$. Furthermore, $P_u > 32$ values are needed to achieve better uniformity, at the cost of further reduction of the FOV. This shows the power of our proposed algorithm which allows high illumination even using $P_u = 4$, without compromising on the FOV.

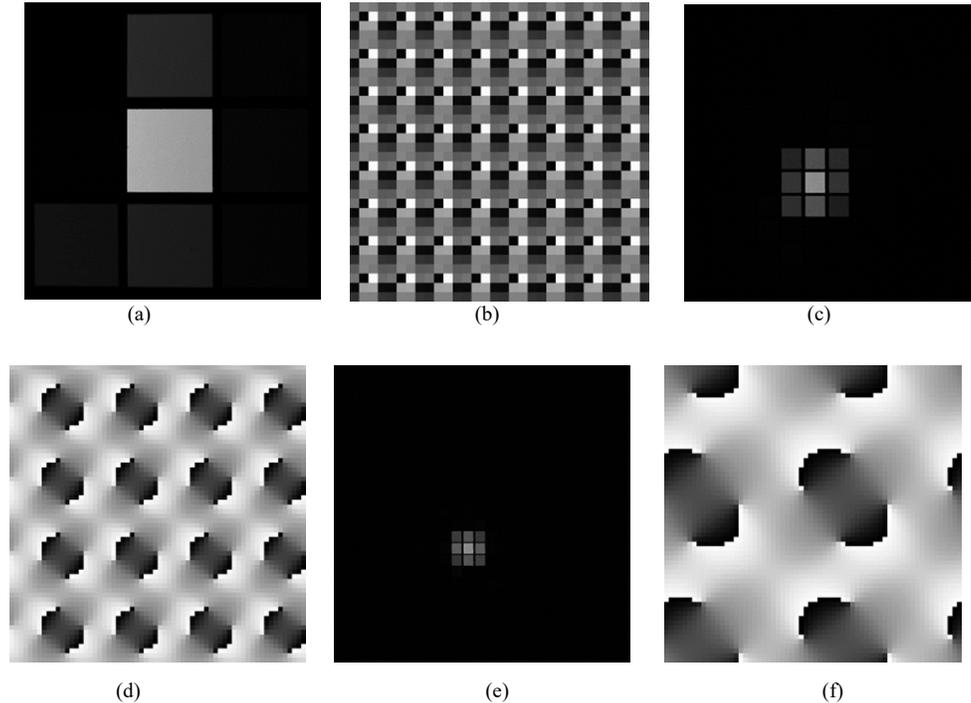

Fig. A8. In Brightfield Imaging mode, the plots show a comparison of chosen Pixflipper algorithm output patterns for each $P_u = 4$, $P_u = 16$ and $P_u = 32$. (a) The camera image resulting from a $P_u = 4$ Pixflipper optimized pattern, (b) a zoomed in 64x64 pixel$^2$ region of the $P_u = 4$ resulting pattern displayed on the SLM, (c) the camera image resulting from a $P_u = 16$ pixflipper optimized pattern, (d) a zoomed in 64x64 pixel$^2$ region of the $P_u = 16$ resulting pattern displayed on the SLM, (e) the camera image resulting from a $P_u = 32$ pixflipper optimized pattern, and (f) a zoomed in 64x64 pixel$^2$ region of the $P_u = 32$ resulting optimized pattern displayed on the SLM.

### 6.6. In situ iterative calibration routine implementation on a different SLM

To demonstrate the universality of our algorithm across SLMs, the Holoeye Pluto-VIS-056 SLM is replaced with a Hamamatsu X10468-07 LCOS-SLM. $P_u = 4$ is used. The X10468 has a larger pixel pitch of 20 μm and has a pixel resolution of 800 × 600 pixels. The Brightfield imaging mode is deployed, and the same algorithm parameters, including $P_u = 4$, is used for this demonstration with the results summarized in Fig. A9. Fig. A9(a) is an example output image due to a Pixelflipper optimized SLM pattern. Note that due to the 20 μm pixel pitch value of this SLM versus the 8 μm of the SLM used earlier, the diffraction angle is smaller bringing the orders closer together on the imaging sensor; therefore the Rectangular Aperture in the optical path is adjusted to prevent the FOVs of the subimages from overlapping with each other. Fig. A9(b) is an output image due to our algorithm optimized pattern, showing a clear increase in the illumination uniformity across the central 3 × 3 orders, in comparison to Fig. A9(a). Fig. A9(c) shows boxplots of $M_{BF}$ values due to the Pixelflipper (90 iterations) and our in situ iterative calibration routine (40 iterations) algorithms, demonstrating the superior performance and applicability of the algorithm. Fig. A9 illustrates the effectiveness of our routine in overcoming the hardware related issues of the Hamamatsu SLM to provide near-uniform intensity spread across the multifocal subimages.

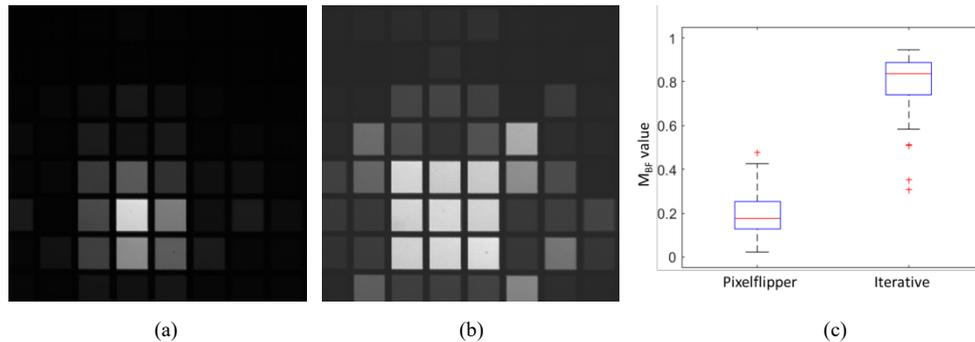

(a)          (b)          (c)

Fig. A9. Optimization results after replacing the Holoeye Pluto-VIS-056 SLM with the Hamamatsu X10468-07 SLM using $P_u = 4$. (a) output image due to a chosen Pixelflipper algorithm optimized SLM pattern, (b) output image due to our algorithm optimized SLM pattern, and (c) boxplots of M values resulting from 90 iterations of the Pixelflipper and 40 iterations of the our algorithm output. The plot in (c) demonstrates the superior illumination intensity distribution performance of our in situ iterative algorithm.

### 6.7. Wavelength dependence of in situ iteratively optimized patterns

The calibration routine patterns optimized for a specific wavelength band are empirically found to give different intensity distributions in the subimages at a different wavelength band. This is demonstrated in Brightfield imaging mode and illustrated in Fig. A10. Fig. A10(a) shows the image resulting from a 685 nm centered bandpass filter with a pattern optimized for this wavelength using our method. When the emission filter is changed to be centered at 510 nm with a bandwidth of 15 nm, the resulting image acquired is shown in Fig. A10(b) using the same pattern as used for Fig. A10(a). Fig. A10(b) shows an undesirable intensity distribution among the subimages. Note that whenever the emission filter is changed, the Rectangular Aperture is adjusted to prevent the FOVs of the subimages from overlapping. For SLM operations, it is recommended to use an updated calibration curve when switching to different wavelengths, therefore as a next step, the calibration settings are updated to give a 2π phase range for 532 nm, which is close to 510 nm for the purpose of this demonstration. In this updated setting, the same pattern which is optimized for a 685 nm bandpass filter is displayed on the SLM and the resulting image acquired and shown in Fig. A10(c), still showcasing a far from ideal intensity spread. Finally, the algorithm is executed using the 532 nm based updated SLM calibration settings and deploying the 510 nm centered emission filter, and the image resulting from this optimized output pattern is shown in Fig. A10(d) which represents a much more uniform intensity distribution as compared to Fig. A10(b) and Fig. A10(c). It is recommended to deploy our optimization routine to separately acquire ideal SLM patterns for each wavelength band desired in the multifocal microscope.

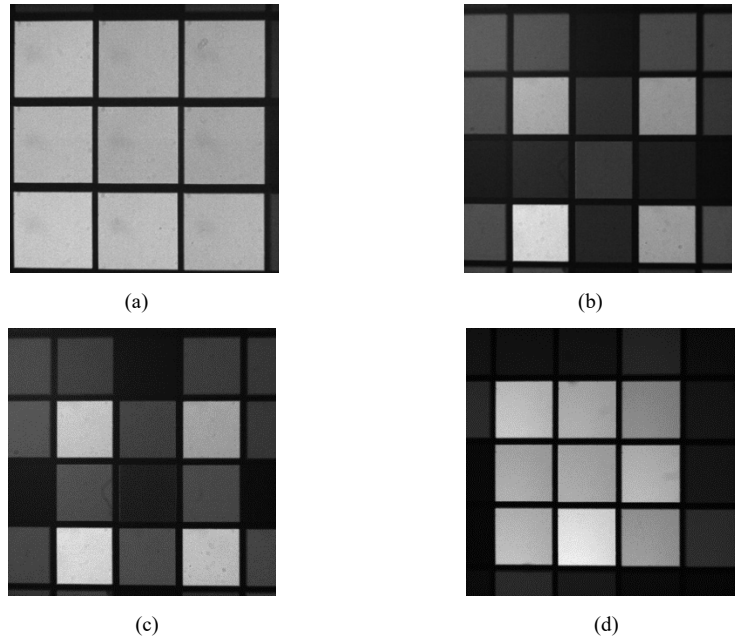

Fig. A10. *Demonstration of wavelength dependence of the iterative optimization algorithm. (a) image resulting from optimized for 685 nm centered emission filter, acquired using 685 nm centered emission filter, with SLM calibration settings suited to 685 nm, (b) image resulting from our algorithm optimized for 685 nm centered emission filter, acquired using 510 nm centered emission filter, with SLM calibration settings suited to 685 nm, (c) image resulting from optimized for 685 nm centered emission filter, acquired using 510 nm centered emission filter, with SLM calibration settings suited to 510 nm, and (d) image resulting from in situ iteratively optimized for 510 nm centered emission filter, acquired using 510 nm centered emission filter, with SLM calibration settings suited to 510 nm.*

## Funding

This work was supported through an Eric and Wendy Schmidt Transformative Technology Fund award to H.Y., S.P., and J.W.S. and by the National Science Foundation, through the Center for the Physics of Biological Function (PHY-1734030).

## Acknowledgements

We would like to thank Matthew King (Princeton University) for preparation of the cell samples.